\documentclass[twocolumn,secnumarabic,amssymb, aps, prl]{revtex4}

\renewcommand{\andname}{\ignorespaces}

\usepackage{graphicx}
\usepackage[utf8]{inputenc}
\usepackage[english]{babel}
\usepackage{textcomp}
\usepackage{natbib,hyperref}
\usepackage{color}
\usepackage{siunitx}
\usepackage{natbib}
\begin{document}

\newcommand{\com}[1]{\textcolor{red}{(#1)}}
\title{Observation of intermolecular Coulombic decay in liquid water}%

\author{Pengju Zhang$^{1}$\footnote{These authors equally contributed to this work.}}
\email[]{pengju.zhang@phys.chem.ethz.ch}
\author{Conaill Perry$^{1*}$}
\author{Tran Trung Luu$^{1,2}$}
\author{Danylo Matselyukh$^{1}$}
\author{Hans Jakob W{\"o}rner$^{1,}$}
\email[]{hwoerner@ethz.ch}
\affiliation{$^{1}$Laboratory for Physical Chemistry, ETH Z{\"u}rich, Vladimir-Prelog-Weg 2, 8093 Z{\"urich}, Switzerland}

\affiliation{$^2$Department of Physics, The University of Hong Kong, Pokfulam Road, SAR Hong Kong, People's Republic of China}
\date{\today}

\begin{abstract}

 We report the first observation of intermolecular Coulombic decay (ICD) in liquid water following inner-valence ionization. By combining a monochromatized table-top high-harmonic source with a liquid micro-jet, we recorded electron-electron coincidence spectra at two photon energies that identify the ICD electrons, together with the photoelectrons originating from the 2a$_1$ inner-valence band of liquid water. Our results confirm the importance of ICD as a source of low-energy electrons in bulk liquid water and provide quantitative results for modeling the velocity distribution of the slow electrons that are thought to dominate radiation damage in aqueous environments.
\end{abstract}
\maketitle

\begin{figure}[h!]
\begin{center}
\includegraphics[width=3.8in]{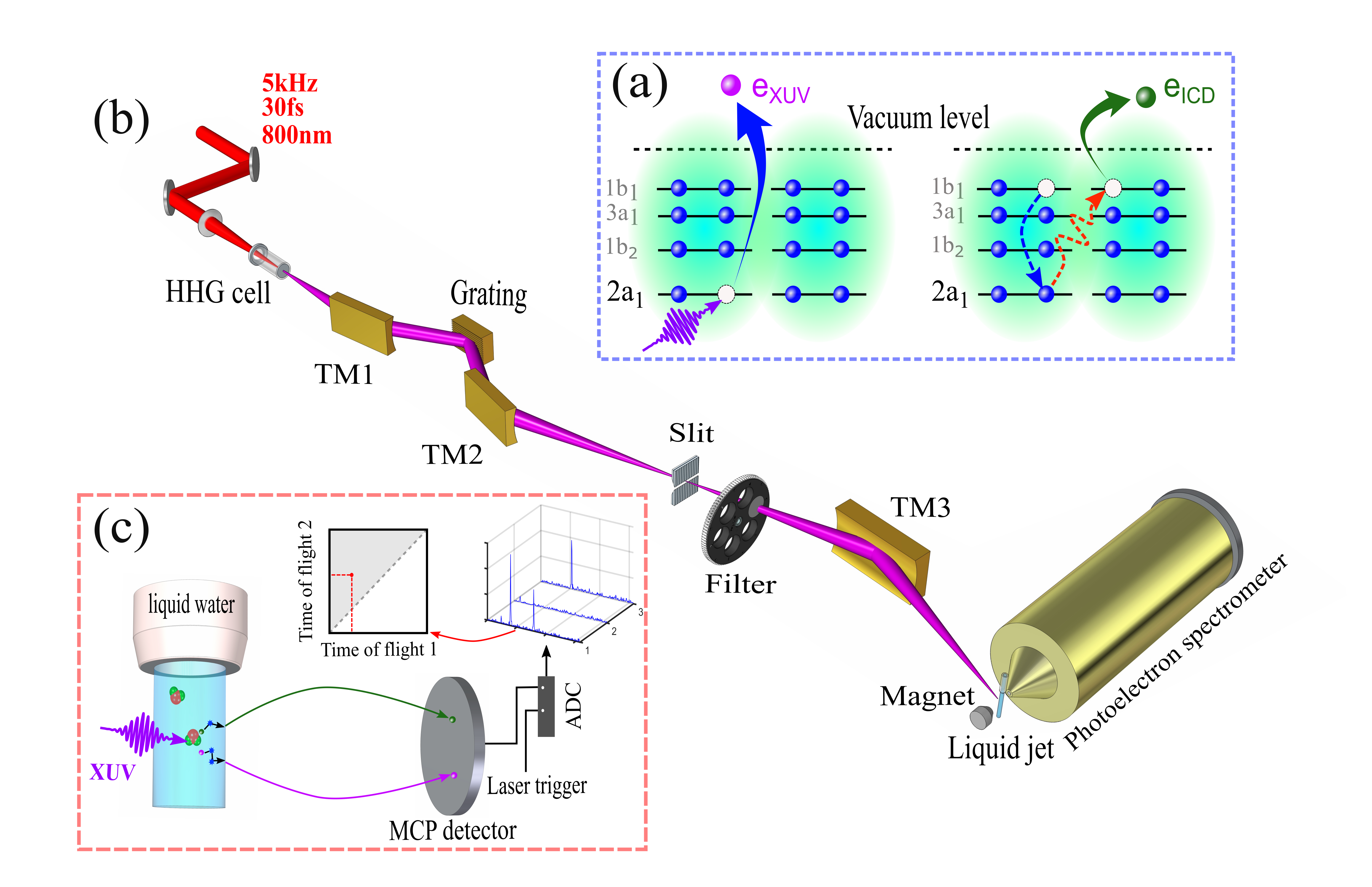}
\end{center}
\vspace*{-3mm}
\caption{(Color online) (a) Schematic representation of the ICD process initiated by inner-valence ionization of ${2a_1}$ orbital in water dimer. The inner-valence vacancy is created by the absorption of XUV photon, an outer-valence electron fills up the inner-valence vacancy via releasing the extra energy to the neighboring water molecule causing a further ionization in its outer-valence orbital. (b) The experimental setup for the photoelectron measurement. (c) Schematic diagram of data acquisition and analysis (See text for the details).}
\label{fig1}
\end{figure}

Charge- and energy-transfer processes are the primary quantum-mechanical events underlying most physical, chemical and biological processes. Inner-valence vacancies created by high-energy incident projectiles (X-rays, $\gamma$-rays, charged particles) are usually filled by one of the valence electrons, whereby the excess energy serves to ionize another valence electron. The most common among such processes is Auger decay, which is energetically possible provided that the inner-shell vacancy lies energetically above the double-ionization threshold. Since this condition is usually not fulfilled for inner-valence ionization, these vacancies relax through internal conversion or fluorescence. If the molecule is embedded in an environment, a faster relaxation channel opens up. In this non-local process, a vacancy in an inner shell of species A relaxes by transferring energy to a neighboring species B, which is ionized by ejection of an electron from its outer-valence shells (Fig.~\ref{fig1}(a)). 

As a ubiquitous, electron-correlation-driven relaxation process, this so called interatomic or intermolecular Coulombic decay (ICD) plays an essential role in understanding the dynamics of energy transfer in weakly-bound complexes and liquids following inner-valence or core-level ionization. Since the pioneering theoretical work of Cederbaum et al. \cite{Cederbaum1997}, great efforts have been put into experimental investigations aiming to identify the existence of ICD in numerous systems, mostly van-der-Waals clusters \cite{Marburger2003,Jahnke2004,Jahnke2007,Morishita2008,Sakai2011,Kim2013,Schnorr2013,Iskandar2015,Yan2018,Hans2020}, nanodroplets \cite{Shcherbinin2017,Kazandjian2018,LaForge2019}, 
water clusters \cite{Jahnke2010,Mucke2010,Richter2018}, hydrated biomolecules \cite{Ren2018} and solids \cite{Wilhelm2017}. Most of the above-mentioned studies have been performed by utilizing synchrotron radiation or free-electron lasers  \cite{Marburger2003,Jahnke2004,Jahnke2007,Morishita2008,Jahnke2010,Mucke2010,Sakai2011,Schnorr2013,Richter2018}. These studies suggested that ICD could be highly relevant in the context of radiation damage because it is an efficient source of slow (0\textendash10 eV) electrons, which are known to be particularly damaging towards bio-organic matter \cite{Bouda2000,Alizadeh2012,Alizadeh2015}. For reviews of this field see Refs. \cite{Jahnke2015, Jahnke2020}.

In living tissues, radiation damage occurs within an aqueous environment, therefore the investigation of ICD in liquid water is important. With the help of synchrotron radiation, ICD-like and related proton-transfer processes have been discussed and partially identified in aqueous solutions following core-level (oxygen 1s) ionization by X-ray radiation \cite{Aziz2008,Thuermer2013,Slavicek2014}. In contrast, ICD following inner-valence ionization in a bulk liquid environment has remained elusive.

Inner-valence ionization of water plays a particularly important role in the cascade of decay processes of high-energy particles, and becomes dominant once their kinetic energies have fallen below the oxygen K-edge as a consequence of inelastic collisions. Therefore, the observation and characterization of ICD in liquid water following inner-valence ionization represent important steps in photochemistry and radiation chemistry. 

The energy distribution of ICD electrons from bulk liquid water is of particular interest for both understanding the radiation-induced mechanisms causing cancer and for their targeted use in cancer therapies. Slow electrons with kinetic energies below $\sim$5~eV are indeed known to only cause single-strand breaks in DNA, whereas faster electrons can also cause double-strand breaks. The latter are much more damaging than the former because they cannot efficiently be repaired \cite{Alizadeh2015}.

In this letter, we report the first observation of inner-valence ICD in liquid water by combining a table-top monochromatized source of high-harmonic radiation, a liquid micro-jet and an electron-electron coincidence photoelectron spectrometer. The experimental setup is shown in Fig.~\ref{fig1}(b). The extreme-ultraviolet (XUV) pulses are provided by a time-preserving monochromator \cite{Conta2016}. High-harmonic generation is driven by a near-infrared laser pulse of $\sim$ 1.2 mJ and $\sim$\ 30~fs duration centered at 800~nm with a repetition rate of 5~kHz. The driving pulse is focused into a semi-infinite gas cell filled with 25~mbar neon. The generated harmonics are collimated by a toroidal mirror (TM1) and spatially diffracted by a \SI{600} {lines/\mm} grating mounted in conical-diffraction geometry. The separated harmonics are further focused by a second toroidal mirror (TM2) onto a 70 $\mu$m slit to select a single harmonic order. Finally, the selected harmonic is imaged by a third toroidal mirror (TM3) onto the liquid microjet in the interaction chamber. 
Liquid water is delivered into the chamber through a 25~$\mu$m inner-diameter quartz nozzle, which is capped with Cu tape, held together by Sn solder to prevent the insulating quartz from charging up due to stray electrons. The electrokinetic charging effect of the jet is minimized by adding NaCl at a concentration of 50~mmol/L. Moreover, a bias voltage of 0.45~V is applied to the liquid jet to simultaneously compensate the effects of the residual streaming potential and that of the vacuum-level offset between the jet and the photoelectron spectrometer \cite{Perry2020}.
The electrons emitted from the liquid jet are recorded by a magnetic-bottle photoelectron spectrometer, previously described in \cite{jordan15a}, consisting of a permanent magnet (1 T), holding a conical iron tip, a 910~mm-long flight tube equipped with a solenoid that generates a homogeneous magnetic field of 1 mT along the flight tube. In this configuration, electrons with a pitch angle between 0$^{\circ}$ and $\sim$120$^{\circ}$ are collected, corresponding to a solid angle of 2.8$\pi$ sr \cite{jordan17c}. 
The spectrometer is divided into the high-pressure interaction region and the low-pressure flight tube by a graphite-coated skimmer. A pair of micro-channel plates (MCP, Hamamatsu) in Chevron configuration with a single-event response time of 150 ps and 27 mm effective area are installed at the end of the flight tube. In order to enhance the detection efficiency of the slow electrons ($<$ 1~eV), a bias potential (+0.9 V) is applied both on the skimmer and the flight tube. The photoelectron spectrometer was calibrated under these conditions using argon and xenon gas delivered through an electrically grounded metallic nozzle.
The raw MCP signal is amplified 10 times by a home-built fast pre-amplifier and then recorded by an amplitude-to-digital (ADC) converter. The waveforms are collected under extremely low count-rates ($\sim$\ 0.08-0.09 counts/laser pulse) to better facilitate coincidence measuring conditions. The dead time is determined by the 0.5-ns time resolution of the ADC device. Typical data acquisition times amounted to 135 minutes per data set. The individual signal waveforms were then processed, using an optimized threshold to reduce the contribution from ringing, and sorted into single, double and multi-hit ($\ge$3) event components. The double-hit events were selected and represented as a coincidence map of the corresponding electron pairs (Fig.~\ref{fig1}(c)). Electronic artefacts originating from MCP ringing were suppressed by discarding electron pairs with a time-of-flight difference of less than 12~ns.

\begin{figure}
	\begin{center}
	\includegraphics[width=3.4in]{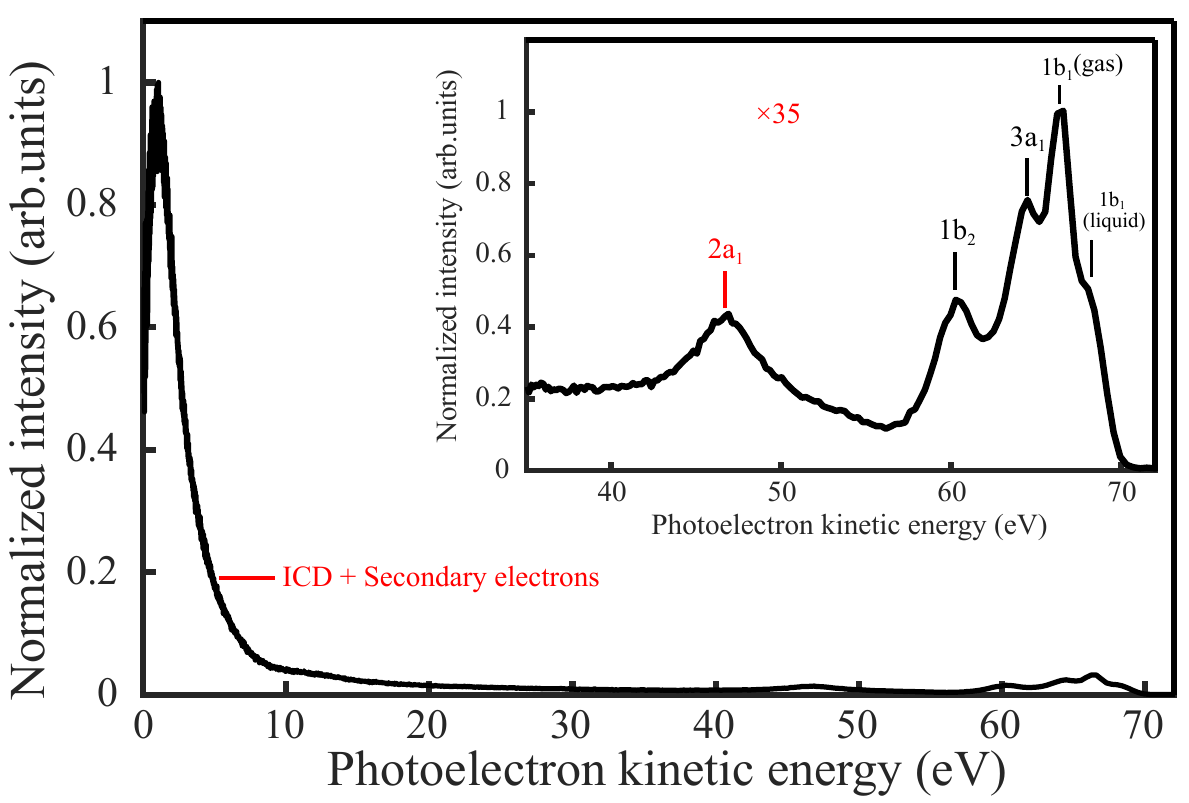}
	\end{center}
	\vspace*{-5mm}
	\caption{(Color online) Photoelectron spectrum of a liquid-water microjet recorded with XUV pulses centered at 79.0~eV (H51 of 800~nm).}
	\label{fig2}
\end{figure}

Figure~\ref{fig2} shows a (non-coincident) photoelectron spectrum of liquid water recorded with a photon energy of 79.0~eV. The signals above kinetic energies of $\sim$58.0~eV originate from the ionization of the outer-valence 1b$_1$, 3a$_1$ and 1b$_2$ bands of liquid water. They are overlapped with the corresponding signals from water vapor that continuously evaporates from the liquid-jet surface. Only the 1b$_1$ photoelectrons from the gas phase are visible as a distinct peak, as highlighted in the inset.
The ionization of the inner-valence (2a$_1$) band of liquid water leads to a very broad photoelectron spectrum extending from $\sim$40 to $\sim$58~eV, in agreement with previous measurements \cite{Winter2004,nishizawa11a,Gadeyne2022}. The slowest electrons between 0 and $\sim$10~eV are dominated by three contributions: i) the so-called secondary electrons that have been produced by electron-impact ionization of water molecules by (primary) photoelectrons \cite{faubel97a}, ii) photoelectrons that have lost a sufficient amount of energy in inelastic collisions, such as electronically inelastic collisions (ionization, excitation, etc.) or many vibrationally inelastic collisions \cite{Gadeyne2022}, and iii) ICD electrons. Separating these contributions requires electron-electron-coincidence spectroscopy.

\begin{figure}
\begin{center}
\includegraphics[width=3.4in]{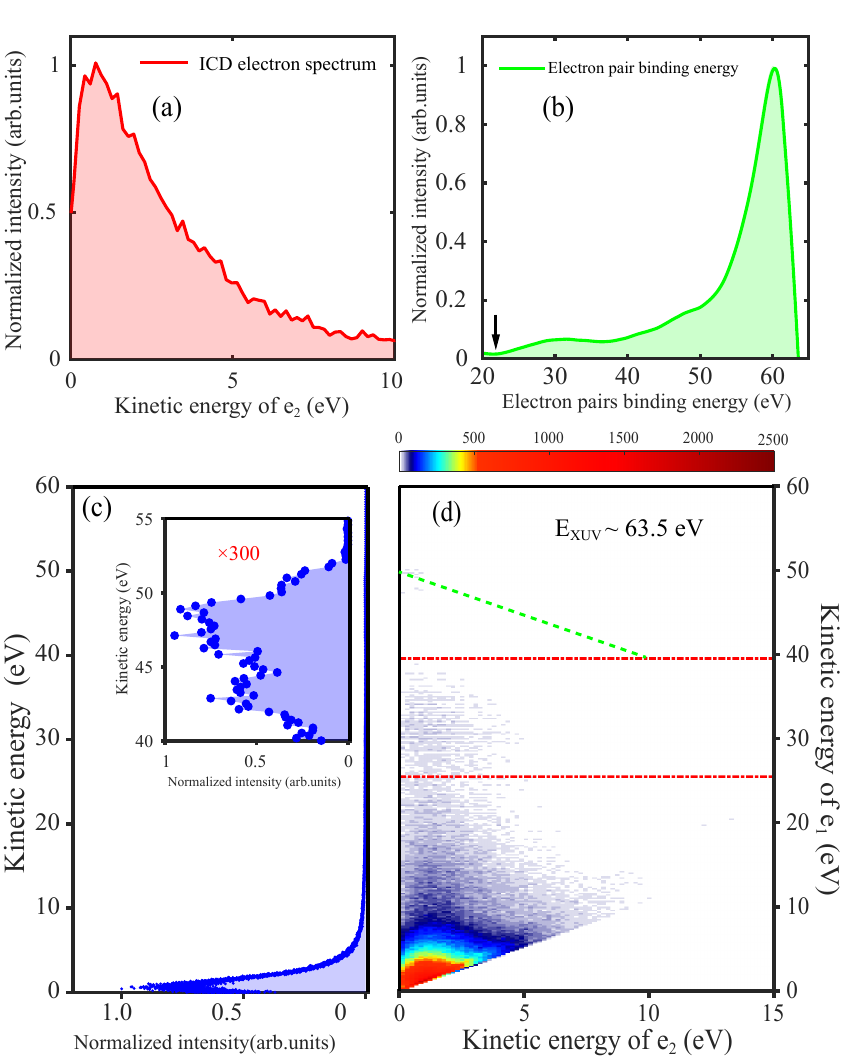}
\end{center}
\vspace*{-5mm}
\caption{(Color online) 
(a) Spectrum of the ICD electrons obtained by vertically summing (d) between the two red
dashed-dotted lines. (b) Distribution of the measured electron pairs as a function of their binding energy, obtained by summing along lines of constant total energy (${E(e_1)}$+${E(e_2)}$), i.e. lines parallel to the green dashed line. (c) Kinetic-energy spectrum of the electron pairs in the double-hit events. The inset is the spectrum ranging from 40~eV to 55~eV with an amplification factor of 300. (d) Coincidence map of electron pairs produced by ionization of a liquid-water microjet with XUV photons of 63.5~eV energy (note the logarithmic intensity scale). The area between the two red lines is dominated by 2a$_1$-photoelectron/ICD-electron pairs.}
\label{fig3}
\end{figure}

Figure~\ref{fig3} shows the electron-electron coincidence spectra of detected electron pairs using an XUV photon energy of 63.5~eV (H41 of 800 nm). Panel (d) shows the coincidence map, i.e. a color-coded histogram of the number of electron pairs corresponding to a given kinetic energy (${e_1}$) of the fast electron, displayed on the vertical axis, and an energy (${e_2}$) of the slow electron, displayed on the horizontal axis. Summing the signals along horizontal lines of the coincidence map yields the "fast-electron" spectrum shown in panel (c), which shows the strong dominance of very slow ($<$5~eV) electrons. 
These slow electrons also dominate the coincidence map shown in panel (d) up to energies of $\sim$10~eV.
Above $e_1\approx$ 10~eV, the coincidence map clearly extends much further along the fast-electron axis, than along the slow-electron axis. The two horizontal dashed red lines identify the region of the ICD electrons, which occur in pairs with photoelectrons from the 2a$_1$ inner-valence band of liquid water. The slow (ICD)-electron distribution is analyzed in more detail below.
The electron pairs located at $e_1<10$~eV, in contrast, are dominated by secondary electrons created by electron-impact ionization and photoelectrons that have lost a large amount of energy, as discussed in detail in our recent work \cite{Gadeyne2022}. The very rare events with $e_1>42$~eV correspond to false coincidences because the simultaneous observation of a "primary" photoelectron with such a high kinetic energy {\it and} a "secondary" electron with any positive kinetic energy electron would violate energy conservation, considering the 63.5~eV photon energy and the 11.67(15)~eV binding energy of liquid water \cite{Perry2020}. The scarcity of these events shows that the count rate ($\sim$0.08 counts/pulse) of the present measurements was sufficiently low to render false coincidences negligible.

Having identified the inner-valence (2a$_1$) ICD signal, we can now determine the characteristic ICD spectrum of liquid water by summing the signals in the coincidence map between the two red dashed-dotted lines, yielding the spectrum shown in Fig.~\ref{fig3}(a). Overall, and in particular for kinetic energies above 2.5~eV, this spectrum displays the typical quasi-exponential shape known from theoretical predictions for small water clusters \cite{mueller06a} and their experimental measurements \cite{Jahnke2010,Mucke2010,Richter2018}.
However, instead of peaking at zero kinetic energy, the ICD spectrum from bulk liquid water exhibits a maximum around 0.8 eV, which indicates the existence of the escape barrier on the surface, as demonstrated in Ref. \cite{Gadeyne2022}. 

Another interesting observable is the distribution function of the electron-pair binding energies. This distribution is obtained by summing the signals of the coincidence map along lines of constant total energy $E(e_1)+E(e_2)$, i.e. along diagonal lines parallel to the dashed green line in  Fig.~\ref{fig3}(d). The obtained distribution function is shown in panel (b). This curve represents all two-hole final states which are populated either by direct double ionization of liquid water, outer-valence (single) photoionization followed by electron-impact ionization or ICD. The black arrow marks the onset of the double-ionization spectrum of liquid water, located at $\sim$23~eV, i.e. much lower than that of isolated water molecules ($\sim$35~eV) \cite{Eland2006}. The onset of the double-ionization spectrum of liquid water lies close to twice the binding energy of the highest-occupied band (1b$_1$) of liquid water \cite{Winter2004,kurahashi2014,Perry2020}, which suggests that this region of low pair-binding energies is associated with the creation of pairs of spatially separated one-hole vacancies (1h1h), rather than double-hole (2h) states. 
The portion of the distribution curve ranging from 23~eV to 38~eV is thus assigned to the production of pairs of 1h1h outer-valence (1b$_1$, 3a$_1$ or 1b$_2$) vacancies. The creation of 2h states is expected to occur in the region of $\sim$38~eV to $\sim$48~eV in isolated molecules \cite{Eland2006}, and therefore slightly lower in liquid water. The higher-binding-energy part of the double-ionization spectrum is dominated by secondary electrons from electron-impact ionization of liquid water by photoelectrons initially emitted from the outer-valence bands and the decelerated photoelectrons themselves \cite{Gadeyne2022}.

\begin{figure}
\begin{center}
\includegraphics[width=3.4in]{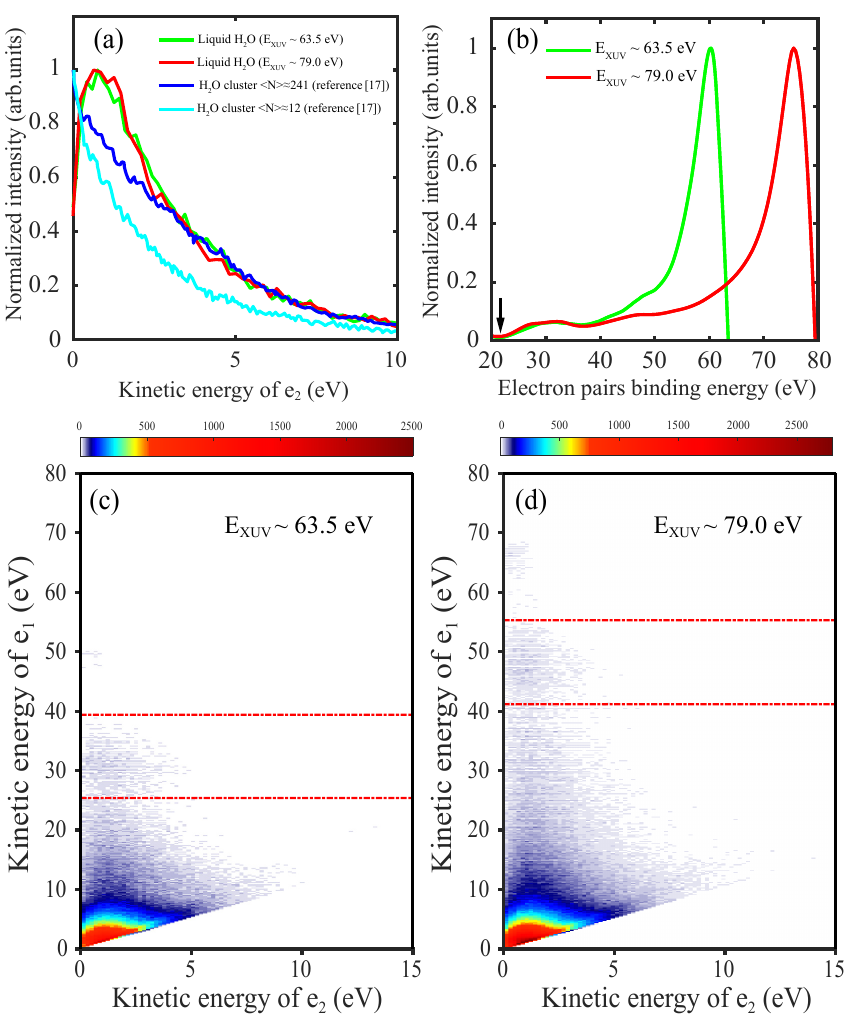}
\end{center}
\vspace*{-5mm}
\caption{(Color online) (a) Comparison of the energy spectra of ICD electrons from water clusters (reference \cite{Richter2018}) and liquid water. The blue and red solid curves are the energy spectra integrated over the rectangle area from panel (c) and panel (d) with the XUV photon energies of 63.5~eV and 79.0~eV, respectively. All of the spectra are normalized to their maximum for comparison. (b) Comparison of the electron-pair spectra as a function of their electron-pair binding energy from panel (c) and panel (d) with the XUV photon energies of 63.5~eV (blue) and 79.0~eV (red), respectively. Panels (c) and (d) represent the coincidence spectra of electron pairs produced by ionization of liquid water with XUV photons of 63.5~eV and 79.0~eV, respectively. The rectangular area in panel (c) covers the energy of ${e_1}$ from 25.5~eV to 39.5~eV, the rectangle area in panel (d) covers the energy of ${e_1}$ from 41.0~eV to 55.0~eV.}
\label{fig4}
\end{figure}

One important further confirmation of the observation of ICD in liquid water consists in varying the XUV photon energy while maintaining the same low count-rate conditions. For the coincidence spectra shown in Fig.~\ref{fig4}(c) and Fig.~\ref{fig4}(d), XUV photon energies of 63.5~eV and 79.0~eV were used, respectively. As expected from our previous discussion, the low-energy part of the coincidence map (${E(e_1),E(e_2)}<10 ~eV$) remains largely unchanged, whereas the extension of the coincidence map along the fast-electron axis is greatly increased. 
This extension is caused by the fact that the kinetic energy of the 2a$_1$ photoelectron is increased as a consequence of the higher photon energy, whereas the distribution along the slow (ICD) electron axis ($e_2$) remains largely unchanged.

Figure~\ref{fig4}(a) shows the energy spectra of ICD electrons from liquid water and their comparison with those obtained from water clusters \cite{Richter2018}. As could be expected, the energy distributions of ICD electrons obtained with 63.5~eV and 79.0~eV XUV photons are almost indistinguishable within their respective signal-to-noise ratios. More interestingly, for kinetic energies above 3~eV, the electron spectra agree well with those obtained from relatively large water clusters, $\langle N\rangle \approx 241$, but are notably broader than those obtained from small water clusters, $\langle N\rangle\approx 12$. 
Below kinetic energies of 3~eV, the liquid-water ICD spectra display a steeper rise, followed by a sharp drop towards 0~eV. This shape is characteristic of the low-energy electron distributions obtained from liquid water in several recent works (\cite{Gadeyne2022} and references therein). In Ref.~\cite{Gadeyne2022}, we have shown that inelastic electron scattering in liquid water leads to the accumulation of electrons at low kinetic energies and that the presence of an escape barrier on the order of 0.2~eV explains the observed drop of the photoelectron signal towards 0~eV. We therefore propose that inelastic electron scattering and the presence of an escape barrier also influence the ICD spectra of liquid water reported in this work and that these two effects partially account for the difference in the ICD spectra of large water clusters and bulk liquid water.

In the absence of inelastic scattering and escape-barrier effects, the kinetic energy of the ICD electron reflects the energy difference between the initially created 2a$_1$ vacancy and the final 1h1h states.
The similarity between the high-energy tails of the ICD spectra of large clusters and bulk liquid water thus suggests that the distribution of energy differences between the 2a$_1$ single-hole and the final 1h1h states are comparable in the two species. This might indicate that the spatially separated 1h1h states created by ICD have a relatively local character, which is consistent with the rapid decay of the ICD rate with the distance $r$ between the two centers ($r^{-6}$ in simple models that neglect orbital overlap).

In conclusion, we have reported the first observation of ICD in liquid water following inner-valence ionization. By combining a table-top XUV light source and a liquid microjet, electron-electron coincidence spectra at two photon energies have been recorded. A clear signature of ICD, i.e. correlated electron pairs consisting of a photoelectron from the 2a$_1$ band and a slow electron have been observed and characterized in detail. The energy distributions of ICD electrons agree very well with those from relatively large water clusters, $\langle N\rangle\approx 241$, but are notably broader than those of small clusters, $\langle N\rangle\approx 12$. This suggests that the distribution of energy differences between the 2a$_1$ single-hole and the final 1h1h states in large water clusters is similar to bulk liquid water and is considerably broader than in small clusters.
Our work provides thus the first quantitative energy distributions of ICD electrons from bulk liquid water. This knowledge will be important for modelling radiation damage in living tissues and evaluating their suitability for radiation-based cancer therapies. Looking forward, our use of an HHG source also represents the first step towards a time-resolved measurement of the time scale of ICD in liquid water. An indirect approach has previously suggested this time to lie between 12~fs and 52~fs in small water clusters \cite{Richter2018}. Calculations on the tetrahedrally coordinated water pentamer \cite{Richter2018} and hydrogen-bonded trimers of the isoelectronic HF molecule \cite{Santra2001,Ghosh2013} suggest significantly shorter lifetimes of 3.9~fs and 3.6-9.8~fs, respectively. Our present experimental observations and methods open a pathway to the direct measurement of the ICD lifetime in liquid water using either terahertz streaking \cite{fruhling15a} on the tens-of-femtosecond timescale or attosecond interferometry \cite{Jordan2015,Jordan20a}, alternatively attosecond streaking \cite{jain17a}, on the few- to subfemtosecond time scale.

\vspace*{3mm}
The authors thank Dr. Matthieu Gisselbrecht (Lund) and Dr. Zhong Yin (Z{\"u}rich) for helpful discussions. This work was supported by the European Research Council (project no. 772797-ATTOLIQ), the Swiss National Science Foundation through the NCCR-MUST project no. 200021\_172946 and ETH Career Seed Grant No SEED-12 19-1/1-004952-000.

\bibliography{ICD,attobib}

\end{document}